\documentclass[prl, reprint, showpacs, superscriptaddress]{revtex4-1}
\usepackage{amsmath, graphicx,amssymb,color, bm}
\begin{document}
\title{Comment on ``Elastic Membrane Deformations Govern Interleaflet Coupling of Lipid-Ordered Domains''}

\date{\today}

\author{J.~J.~Williamson}
\email{jjw@fastmail.com}
\altaffiliation{Current address:\ The Francis Crick Institute, Lincoln's Inn Fields Laboratories, 44 Lincoln's Inn Fields, London WC2A 3LY, UK}
\author{P.~D.~Olmsted}
\email{pdo7@georgetown.edu}
\affiliation{Department of Physics, Institute for Soft Matter Synthesis and Metrology, Georgetown University, 37th and O Streets, N.W., Washington, D.C. 20057, USA}

\maketitle

\begin{figure}
   \includegraphics[width=8.5cm]{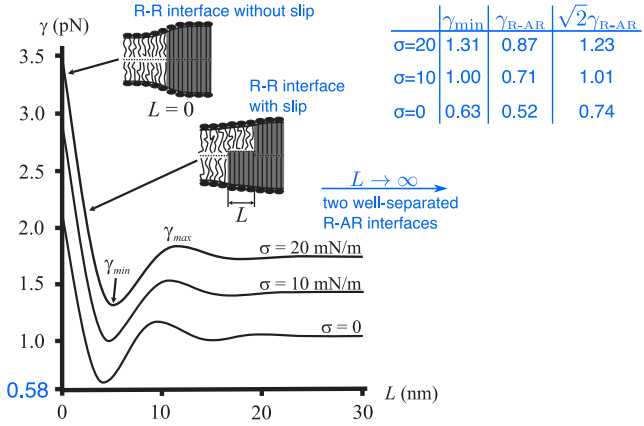} % requires the graphicx package
   \caption{Line tension versus slip length $L$ \cite{Galimzyanov2015}, annotated (blue). $\gamma(L\rightarrow\infty)\!=\!2\gamma_{\textrm{R-AR}}$ (see text). An axis tick is corrected.}
   \label{fig:Galim}
\end{figure}

\noindent Galimzyanov et al.\ \cite{Galimzyanov2015} find that line tension between thick liquid-ordered ($L_{o}$) and thinner liquid-disordered ($L_{d}$) registered lipid bilayer phases is minimised by an asymmetric ``slip region'', length $L\!\sim\!5\,\textrm{nm}$ (Fig.~\ref{fig:Galim}). They claim that line tensions alone explain domain registration, without ``direct'' (area-dependent) inter-leaflet interaction \cite{May2009, Putzel2008}. We show this is unfounded;\ without direct interaction their results would predict \textit{antiregistration}, dependent on composition. 
To find equilibrium from line energies, line \textit{tensions} must be combined with interfacial lengths for given states at given composition. This was not done in \cite{Galimzyanov2015}. 
\color{black}

\begin{figure}
\includegraphics[width=8.5cm]{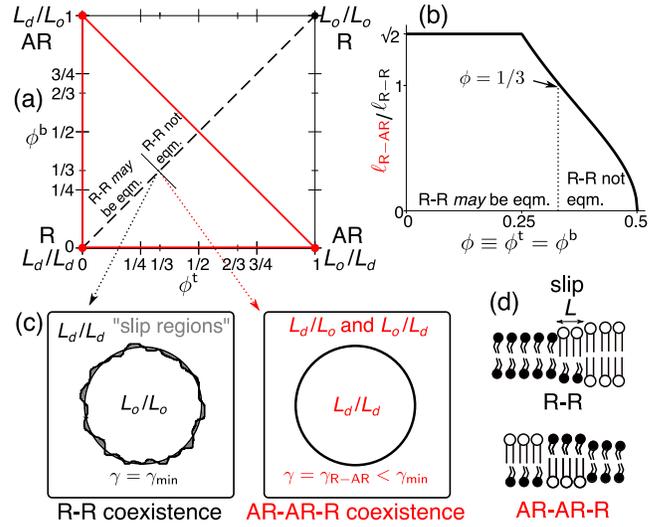}
\caption{\label{figure}(a)~Partial diagram of two-phase R-R (black) and three-phase AR-AR-R (red) coexistences for $\phi^\textrm{b}\!=\!\phi^\textrm{t}\!\equiv\!\phi\!<\!1/2$ {[the full diagram is symmetric under inversion through $(0.5,0.5)$ \cite{Williamson2014, *Williamson2015a}]}. (b)~Minimal R-AR or R-R interface length, comparing AR-AR-R and R-R coexistence. (c)~Morphology at $\phi\!=\!1/3$, beyond which AR-AR-R \textit{decreases} thickness-mismatched interface length. (d)~R-R coexistence;\ AR-AR-R.
}
\end{figure}

Fig.~\ref{figure}a shows ``registered'' (R) and ``antiregistered'' (AR) bilayer phases in $(\phi^\textrm{t},\phi^\textrm{b})$ (top/bottom-leaflet composition) space \cite{May2009, Putzel2008, Williamson2014, *Williamson2015a}. 
We assume $\phi^\textrm{b}\!=\!\phi^\textrm{t}\!\equiv\!\phi$ in each leaflet.
``Domain registration'' is R-R coexistence (Fig.~\ref{figure}a), with slip to minimise line tension (Fig.~\ref{fig:Galim}) \cite{Galimzyanov2015}.

Without direct inter-leaflet coupling, R and AR bulk free energies would be equal \cite{Putzel2008} so line energies alone would determine equilibrium. The authors claim perfect \textit{antiregistration} (AR-AR state) is only possible for $\phi\!=\!1/2$. They assume imperfect antiregistration involves \textit{both} R-AR interfaces in Fig.~\ref{fig:Galim}. They 
find a minimum line tension $\gamma_{\textrm{min}}$, and a limiting value $\gamma_{\infty}\!\equiv\!\gamma(L\rightarrow\infty)\!>\!\gamma_{\textrm{min}}$ as the two R-AR interfaces in Fig.~\ref{fig:Galim} ($L_d$/$L_d$ to AR, AR to $L_o$/$L_o$) move apart. They claim that  $\gamma_{\infty}\!>\!\gamma_\textrm{min}$ favours an R-R state. However, $\phi\!\neq\!1/2$ only forces one R phase (AR-AR-R state, Fig.\ \ref{figure}a), and the line tension of \textit{one} isolated R-AR interface in Fig.~\ref{fig:Galim} is $\gamma_{\scriptscriptstyle \textrm{R-AR}}\!\approx\!\gamma_{\infty}/2$.

In R-R, an $L_{o}$/$L_{o}$ droplet, area $\phi A$, circumference $\ell_{\scriptscriptstyle\textrm{R-R}}\!=\!2\sqrt{\pi\phi A}$ is surrounded by $L_{d}$/$L_{d}$ (Fig.~\ref{figure}c, $A\!=\!\textrm{total area}$). The line energy is $W_{\scriptscriptstyle\textrm{R-R}}\!=\!\ell_{\scriptscriptstyle\textrm{R-R}}\gamma_\textrm{min}$.
For AR-AR-R, the area of $L_o$/$L_d$ or $L_d$/$L_o$ is $2\phi A$. [The equal-thickness phases are treated as a quasi-uniform single phase.] Hence, for $\phi\!<\!1/4$, $L_d$/$L_d$ surrounds an AR domain ($\ell_{\scriptscriptstyle\textrm{R-AR}}\!=\!2\sqrt{ 2 \pi \phi A}$), with $W_{\scriptscriptstyle\textrm{R-AR}}\!=\!{\ell_{\scriptscriptstyle\textrm{R-AR}}}{\gamma_{\scriptscriptstyle\textrm{R-AR}}}$. $\ell_{\scriptscriptstyle\textrm{R-AR}}=\sqrt{2}\ell_{\scriptscriptstyle\textrm{R-R}}$ (Fig.~\ref{figure}b), so R-R would be equilibrium ($W_{\scriptscriptstyle\textrm{R-R}}\!<\!W_{\scriptscriptstyle\textrm{R-AR}}$) \textit{if} $\gamma_{\textrm{min}}\!<\!\sqrt{2} \gamma_{\scriptscriptstyle\textrm{R-AR}}$. This holds for $\sigma\!=\!0$, but not $\sigma\!=\!20\,\textrm{mN/m}$ (Fig.~\ref{fig:Galim}), and will depend on other parameters, e.g., the degree of hydrophobic mismatch. 

For $1/4\!<\!\phi\!<\!1/2$, AR surrounds an $L_d$/$L_d$ domain of $\ell_{\scriptscriptstyle\textrm{R-AR}}\!=\!2\sqrt{ \pi(1- 2\phi)A}$. For $1/3\!<\!\phi\!<\!1/2$ (by symmetry $1/2\!<\!\phi\!<\!2/3$ \cite{Williamson2014, *Williamson2015a}), $\ell_{\scriptscriptstyle\textrm{R-AR}}\!<\!\ell_{\scriptscriptstyle\textrm{R-R}}$;\ R-R would not be equilibrium for $\gamma_{\scriptscriptstyle\textrm{R-AR}}\!<\!\gamma_\textrm{min}$. Indeed, as $\ell_{\scriptscriptstyle\textrm{R-AR}}\!\to\!0$ \textit{ever higher} $\gamma_{\scriptscriptstyle\textrm{R-AR}}$ would be needed to stabilise R-R. A full calculation of $\gamma_{\scriptscriptstyle\textrm{R-AR}}$ should use a single R-AR interface. Imposing flat boundary conditions could give an area-dependent cost, or make $\gamma_{\scriptscriptstyle\textrm{R-AR}}$ depend on where the boundary conditions are enforced.

\color{black}
Upon reinstating direct, area-dependent inter-leaflet coupling [undulations (see Reply) are one possible source], the \textit{bulk} free energy of R is lower than AR, explaining equilibrium domain registration over all $\phi$ \cite{May2009, Putzel2008, Williamson2014, *Williamson2015a}. 

{\small John J.~Williamson and Peter D.~Olmsted, Department of Physics, Institute for Soft Matter Synthesis and Metrology, Georgetown University, 37th and O Streets N.W., Washington, D.C.\ 20057, USA}

\bibliography{bibliography}

\end{document}